\documentclass[%
reprint,
amsmath,amssymb,
prb,
]{revtex4-2}

\usepackage{graphicx}
\usepackage{dcolumn}
\usepackage{bm}
\usepackage{hyperref}

\begin{document}

\author{Nikita~S.~Pavlov}
\email{pavlov@iep.uran.ru}
\affiliation{Institute for Electrophysics, Russian Academy of Sciences, Ekaterinburg 620016, Russia}
\affiliation{P.N. Lebedev Physical Institute, Russian Academy of Sciences, Moscow, 119991, Russia}

\author{Igor~R.~Shein}
\affiliation{Institute of Solid State Chemistry, Ekaterinburg, 620108, Russia}

\author{Kirill~S.~Pervakov}
\affiliation{P.N. Lebedev Physical Institute, Russian Academy of Sciences, Moscow, 119991, Russia}

\author{Vladimir~M.~Pudalov}
\affiliation{P.N. Lebedev Physical Institute, Russian Academy of Sciences, Moscow, 119991, Russia}

\author{Igor~A.~Nekrasov}
\affiliation{Institute for Electrophysics, Russian Academy of Sciences, Ekaterinburg 620016, Russia}
\affiliation{P.N. Lebedev Physical Institute, Russian Academy of Sciences, Moscow, 119991, Russia}

\title{Anatomy of the band structure of the newest apparent near-ambient superconductor LuH$_{3-x}$N$_x$}

\date{\today}
	
\begin{abstract}
	 Recently it was claimed that nitrogen-doped lutetium hydride exhibited a near-ambient superconducting transition with a temperature of 294~K at a pressure of only 10~kbar, this pressure being several orders of magnitude lower than previously demonstrated for hydrides under pressure. 
	In this paper, we investigate within DFT+U the electronic structure  of both parent lutetium hydride LuH$_3$ and nitrogen doped lutetium hydride LuH$_{2.75}$N$_{0.25}$. We calculated corresponding bands, density of states and Fermi surfaces.
	It is shown that in the stoichiometric system the Lu-5d states cross the Fermi level while the H-1s states make almost no contribution at the Fermi level.
	However, with nitrogen doping, the N-2p states enter the Fermi level in large quantities and bring together a significant contribution from the H-1s states.
	The presence of N-2p and H-1s states at the Fermi level in a doped compound can facilitate the emergence of superconductivity.
	For instance, nitrogen doping almost doubles the value of DOS at the Fermi level. Simple BCS analysis shows that the nitrogen doping of LuH$_3$ can provide T$_c$ more than 100K and even increase it with further hole doping.
\end{abstract}


\maketitle

\section{Introduction}
The discovery of high temperature superconductivity under high pressure around 100-250 GPa with $T_c$ about 203~K in a hydrogen-containing H$_3$S system~\cite{pap1} gave rise to a flow of experimental and theoretical works (see reviews~\cite{pap2,pap3,pap4,pap5,nekrasov_hydrides,troyan_UFN_2022}). The subsequent discovery of near-room temperature superconductivity in hydrides brings us back to the classical electron-phonon superconducting pairing, very likely consistent within the Bardeen-Cooper-Schriefer (BCS) theory. Remarkably, there appeared many different classes of hydrides under pressure and band structure calculations have shown the ability to predict crystal structures and T$_c$ values for many of those hydrides~\cite{pap6,pap7,pap8,troyan_YH6_2021,semenok_LaYH10_2021,troyan_UFN_2022}.

The stunning experimental discovery of the near room temperature superconductors in the so-called hydrides under high-pressure indeed is the hottest topic in modern physics. 
For potential practical applications of this class of superconductors it is crucial to lower the value of the external pressure at which superconductivity occurs. 
Here we address the issue of superconductivity in the recently discovered nitrogen doped superconducting rare-earth  hydrides.
Recent report on apparent superconductivity in the nitrogen-doped lutetium hydride LuH$_{x}$N$_y$ with $T_c=294$~K at 10 kbar~\cite{LuH3_dasenbrock-gammon_Nature}, has sparked intensive theoretical and experimental research.
Three months later, resistivity measurements on the same samples from~\cite{LuH3_dasenbrock-gammon_Nature} were repeated by another scientific group and show the superconductivity with $T_c=240$~K at 8.5~kbar~\cite{LuH3_Salke_Evidence} and above. However, in Ref.~\cite{LuH3_Salke_Evidence} some samples did not superconduct due to the way of preparation as marked by the authors.
There are many theoretical works on crystal structure of LuH$_{x}$N$_y$, which is most stable at given experimental conditions~\cite{LuH3_hilleke_structure,LuH3_sun_stability_cubic,LuH3_huo_fcc,LuH3_liu_parent_2023,LuH3_ferreira_search_sc,LuH3_xie_phase_diagram}. Also these works report the calculated phonon spectra and provide estimation of possible BCS-type superconductivity.
Strictly speaking, none of these works demonstrate the emergence of BCS-type superconductivity with T$_c$=294K at 10 kbar.
In addition, it has also been suggested that there may exist a more stable LuH$_2$ phase of the CaF$_2$-type (Fm3m) with a very low $T_c = 0.026$~K~\cite{LuH3_lu_electron-phonon}.
In the paper~\cite{LuH3_lucrezi_effects} it was shown that taking into account temperature and quantum anharmonic lattice effects in the phonon calculations for LuH$_3$ stabilizes within the Fm$\bar{3}$m space group at pressure up to 4.4~GPa and temperature of 150~K. The superconductivity with $T_c$ value in the range of $50-60$~K is also obtained in Ref.~\cite{LuH3_lucrezi_effects}.
The non-trivial topology of bands and surface states near the Fermi level of stoichiometric LuH$_3$ was found in Ref.~\cite{LuH3_sufyan_topological}.
The optical spectra of LuH$_2$(Fm$\bar{3}$m), LuH(P$\bar{4}$3m), LuH (F$\bar{4}$3m) and LuN (Fm$\bar{3}$m) were theoretically investigated~\cite{LuH3_tao_leading_components}.
Based on DFT calculations the reflectivity of LuH$_2$(Fm$\bar{3}$m) was calculated to explain the observed colour change under pressure~\cite{LuH3_kim_microscopic}.
Extended theoretical comments on the original paper~\cite{LuH3_dasenbrock-gammon_Nature} were presented in Ref.~\cite{LuH3_hirsch_comment}.
The identified classes of structures and hydrogen-vacancy ordering schemes for Lu-H-N properties was preformed in Ref.~\cite{LuH3_denchfield_flat_2023}.
Ab initio study of the structural, vibrational and optical properties LuH$_2$ and LuH$_3$ were calculated in Ref.~\cite{LuH3_dangic_ab_2023}.

There are also several experimental attempts to reproduce the original results~\cite{LuH3_dasenbrock-gammon_Nature} with the same chemical compositions or newly synthesized samples.
The resistivity and magnetic properties of Lu, LuH$_2$ were measured~\cite{LuH3_zhang_electronic} at temperatures from 300 down to 2 K. Here the resistivity of lutetium dihydride was found to be temperature independent.
The resistivity measurements for both LuH$_2$ and LuH$_3$ showed no evidence of superconductivity~\cite{LuH3_cai_No_evidence} in the temperature range of $300 - 4$~K and pressure range of $0.9 - 3.4$~GPa. Also magnetic susceptibility measurements in the pressure range of $0.8-3.3$~GPa and temperatures down to 100~K did not show any signatures of superconductivity.
The resistivity and specific heat of LuH$_2$ were measured in the paper~\cite{LuH3_wang_percolation}. It is shown that percolation of the metallic grains through the insulating surfaces produces a sharp drop in resistivity.
The resistivity measurements in Ref.~\cite{LuH3_shan_pressure_color} with LuH$_2$ under elevated pressures up to $\sim 7$ GPa did not reveal superconductivity down to 1.5~K.
The nitrogen-doped compound LuH$_{2\pm x}$N$_y$ shows a metallic behavior without superconductivity down to 10~K and pressures from 1~GPa to 6~GPa~\cite{LuH3_ming_absence_sc_pressure}.
Possible structural/electronic phase transition between two non-superconducting phases was observed in LuH$_{2\pm x}$N$_y$~\cite{LuH3_xing_observation_nonsc}, which is most pronounced in the ``pink'' phase and might have been erroneously interpreted as a sign of superconducting transition.
The resistivity of LuH$_{2\pm x}$N$_y$ under pressures up to 50.5~GPa shows progressively optimized metallic behavior with pressure~\cite{LuH3_zhang_pressure_color}.
The reflectivity of LuH$_2$ under pressure was investigated in~\cite{LuH3_zhao_pressure_optic} and was found to change significantly in the visible spectrum due the plasmon resonance. Also the pressure-induced color change in LuH$_{2\pm x}$N$_y$ was observed in Ref.~\cite{LuH3_liu_pressure_color_2023} up to 33~GPa.
Raman spectroscopy and X-ray diffraction were used to characterize the crystal structure of LuH$_3$ at various pressures~\cite{LuH3_moulding_raman_xray}.
Single-crystalline films of LuH$_{2+x}$ were studied in Ref.~\cite{LuH3_li_films} using Raman spectroscopy and electrical transport measurements.
In Ref.~\cite{LuH3_Guo_Magnetism_Pressure} magnetic susceptibility was measured for the sample Lu-N-H pressure up to 4.3~GPa and did not find the superconductivity.
Besides the original claim made in Ref.~\cite{LuH3_dasenbrock-gammon_Nature} and recent Ref.~\cite{LuH3_Salke_Evidence} the other work~\cite{LuH3_li_sc_above70} in which superconductivity with $T_c =71$~K at 218~GPa was found, presumably for the LuH system with the space symmetry group Fm$\bar{3}$m.
With lowering pressure to 181~GPa, T$_c$ decreases down to 65K~\cite{LuH3_li_sc_above70}, thus approaching  the results previously reported in~\cite{LuH3_shao_SC_2021} for LuH$_3$ ($T_c =12.4$~K at 122~GPa).

At this moment we are not aware of any detailed study of band structure of the N-doped lutetium hydride LuH$_{3-x}$N$_x$ (x=0.25) and its relation to that for the parent LuH$_3$ compound.
In this paper we performed DFT+U band structure calculations of both parent lutetium hydride LuH$_3$ and nitrogen doped lutetium hydride LuH$_{2.75}$N$_{0.25}$. We calculated the bands with orbital contributions, density of states and Fermi surfaces.
Within the framework of the Wannier projection method a minimal model which gives good description of the bands crossing the Fermi level is found.
In particular, our analysis reveals the contribution of nitrogen  levels in the  density of states at the Fermi level and in the formation of superconducting state.
We also provide simple BCS-analysis of T$_c$ for materials under consideration.

\section{Computational Details}
The calculations were performed within the GGA+U approximation in the VASP software package~\cite{vasp}. The generalized gradient approximation (GGA) in the form of the Perdew-Burke-Ernzerhof (PBE) exchange-correlation functional~\cite{DFT_PBE} was employed. The strong onsite Coulomb repulsion of Lu-$4f$ electrons was described with the GGA+U scheme with the simplified Dudarev approach~\cite{DFT_U} (U=5.0~eV).

\begin{figure}[h]
	\includegraphics[width=0.75\linewidth]{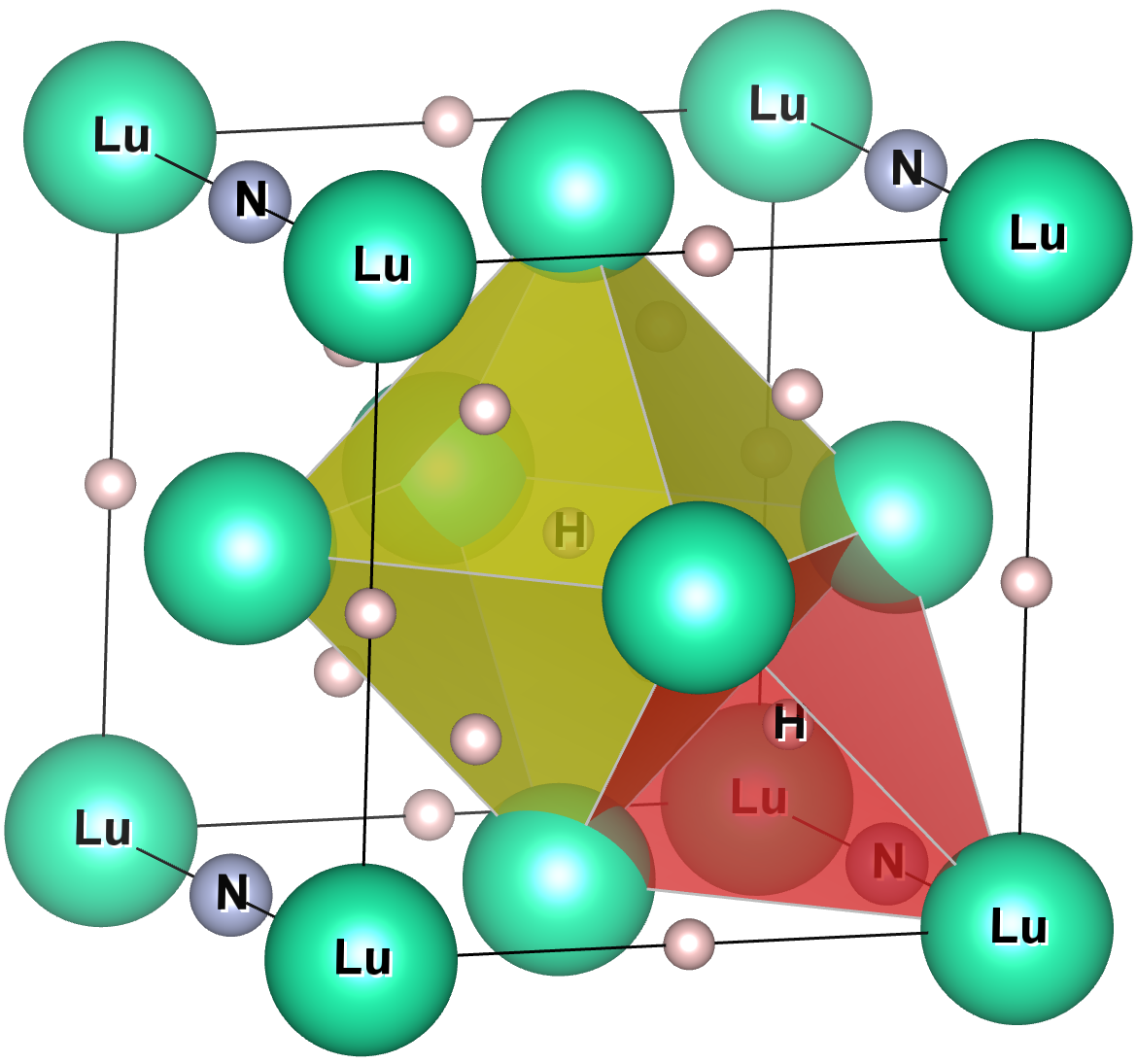}
	\caption{Crystal structure of LuH$_{2.75}$N$_{0.25}$ with two types of H atom surroundings (yellow - octahedral, red - tetrahedral).}
	\label{LuH3_structure}
\end{figure}

We consider the parent lutetium hydride LuH$_3$ and nitrogen doped lutetium hydride LuH$_{2.75}$N$_{0.25}$ with the spatial symmetry group $Fm\bar{3}m$ and lattice parameter $a = 5.0289$~\AA~\cite{LuH3_dasenbrock-gammon_Nature} at ambient pressure. We select this parameters corresponding to the ambient condition, because the band structure and density of states is practically the same with those for parameter $a = 5.007$~\AA~determined experimentally at pressure of 1~GPa~\cite{LuH3_dasenbrock-gammon_Nature}. The H atoms are located in positions with tetrahedral (0.25,0.25,0.25) and octahedral (0.5,0,0) surroundings Lu atoms (Fig.~\ref{LuH3_structure}).
The Lu atoms are located at the point of origin, at (0,0,0).
To get LuH$_{2.75}$N$_{0.25}$ system only one of the four hydrogen atoms at octahedral surrounding is replaced by the nitrogen atom, since, such substitution, as shown in the work~\cite{LuH3_dasenbrock-gammon_Nature}, leads to a metal. The ion relaxation was done for LuH$_{2.75}$N$_{0.25}$.
In Ref.~\cite{LuH3_huo_fcc} it was obtained that the enthalpy of formation of LuH$_{2.75}$N$_{0.25}$ is negative at 1~GPa, which means  that this compound can be formed.
The full unit cell was calculated for ability of comparison of LuH$_3$ and LuH$_{2.75}$N$_{0.25}$.

Wannier functions were obtained using the Wannier90~\cite{Wannier90_Pizzi2020} package with projection onto H-1s in octahedral environment and N-2p orbitals.

\section{Results and Discussion}
\subsection{Band structure and DOS}
\begin{figure}[h]
		\includegraphics[width=0.9\linewidth]{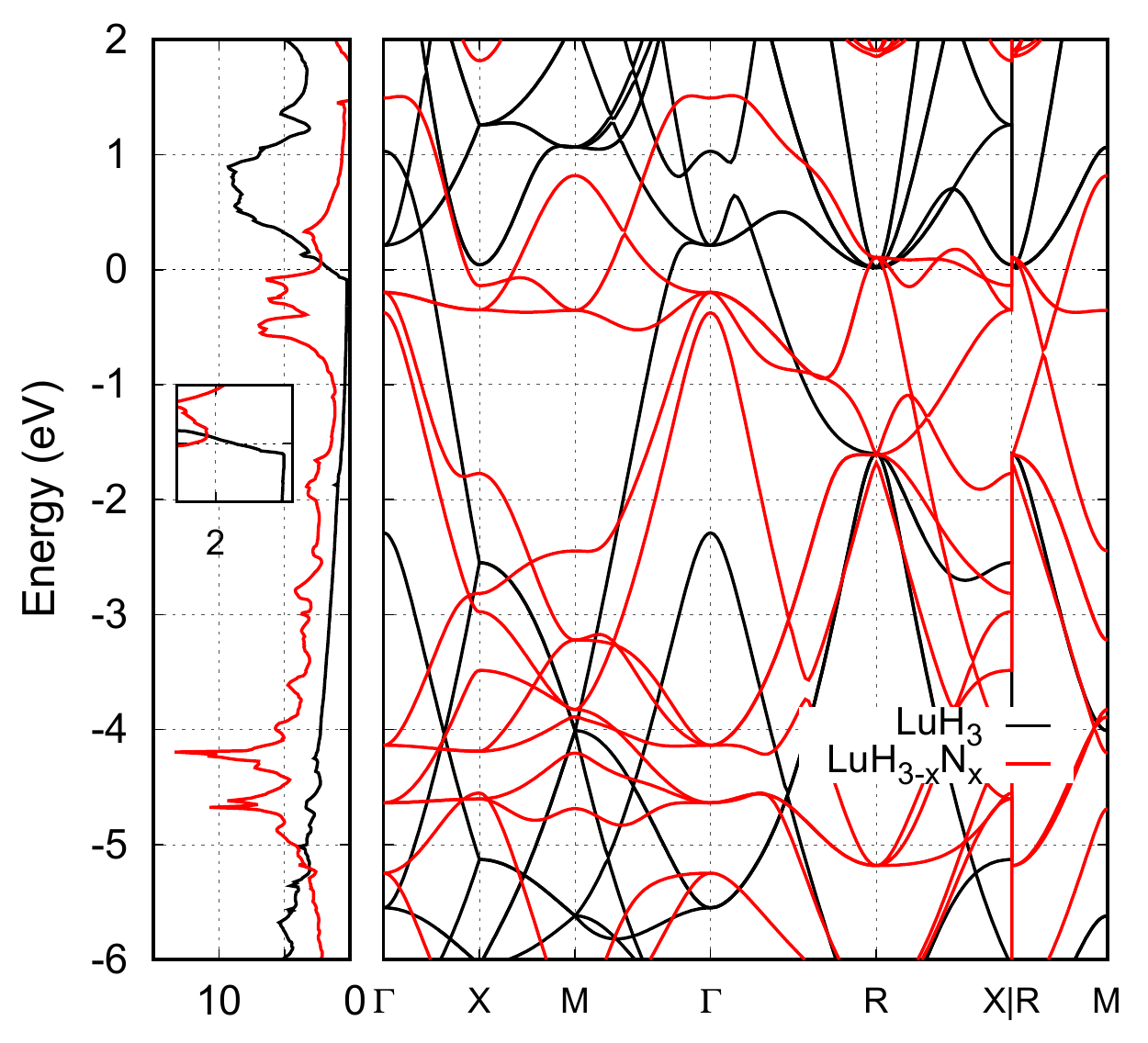}
	\caption{DFT/GGA (left panel) total densities of states and (right panel) band dispersions of LuH$_{3}$ (black lines), LuH$_{2.75}$N$_{0.25}$ (red lines). The inset on left panel shows the densities of states near the Fermi level [-0.5;0.5]~eV. Note that the DOS curves are shifted relative each other to superpose the Fermi levels at zero energy.}
	\label{LDA_DOS}
\end{figure}
The comparison of DFT/GGA total densities of states (left panel) and band dispersions (right panel) between LuH$_{3}$ (black lines) and LuH$_{2.75}$N$_{0.25}$ (red lines) is shown on Fig.~\ref{LDA_DOS}. 

Nitrogen doping leads to formation of a wide peak in total DOS just below the Fermi level
(red line on the left panel of Fig.~\ref{LDA_DOS}),
in contrast to the case of stoichiometric LuH$_{3}$. The most important is that the total density of states at the Fermi level increases almost by a factor of two in LuH$_{2.75}$N$_{0.25}$ from 1.5 to 2.6 states/eV/u.c. (left panel of Fig.~\ref{LDA_DOS}). Later we will use this fact to estimate the possible superconducting T$_c$.

For both compounds we have well pronounced three dimensional bands.
The substitution of hydrogen atom by nitrogen one gives four additional holes per unit cell which leads to the $\sim$1.8~eV lowering of the Fermi level in LuH$_{2.75}$N$_{0.25}$ as compared to LuH$_{3}$. Although the bands can not be superposed by their simple shift, some features are quite similar for both systems.

In the case of LuH$_{2.75}$N$_{0.25}$ there are several bands which cross the Fermi level at each high-symmetry direction, while for LuH$_{3}$ only couple of bands near $\Gamma$-point cross the Fermi level.
For LuH$_{2.75}$N$_{0.25}$ at the Fermi level there are pronounced flat band regions in the vicinity of R and X points, which are missing for LuH$_{3}$. These local flat bands are favourable to superconducting pairing.
Near R-point both compounds have bands nearly touching the Fermi level.

\subsection{Orbital projected bands}
\begin{figure*}[!ht]
	\includegraphics[width=0.32\linewidth]{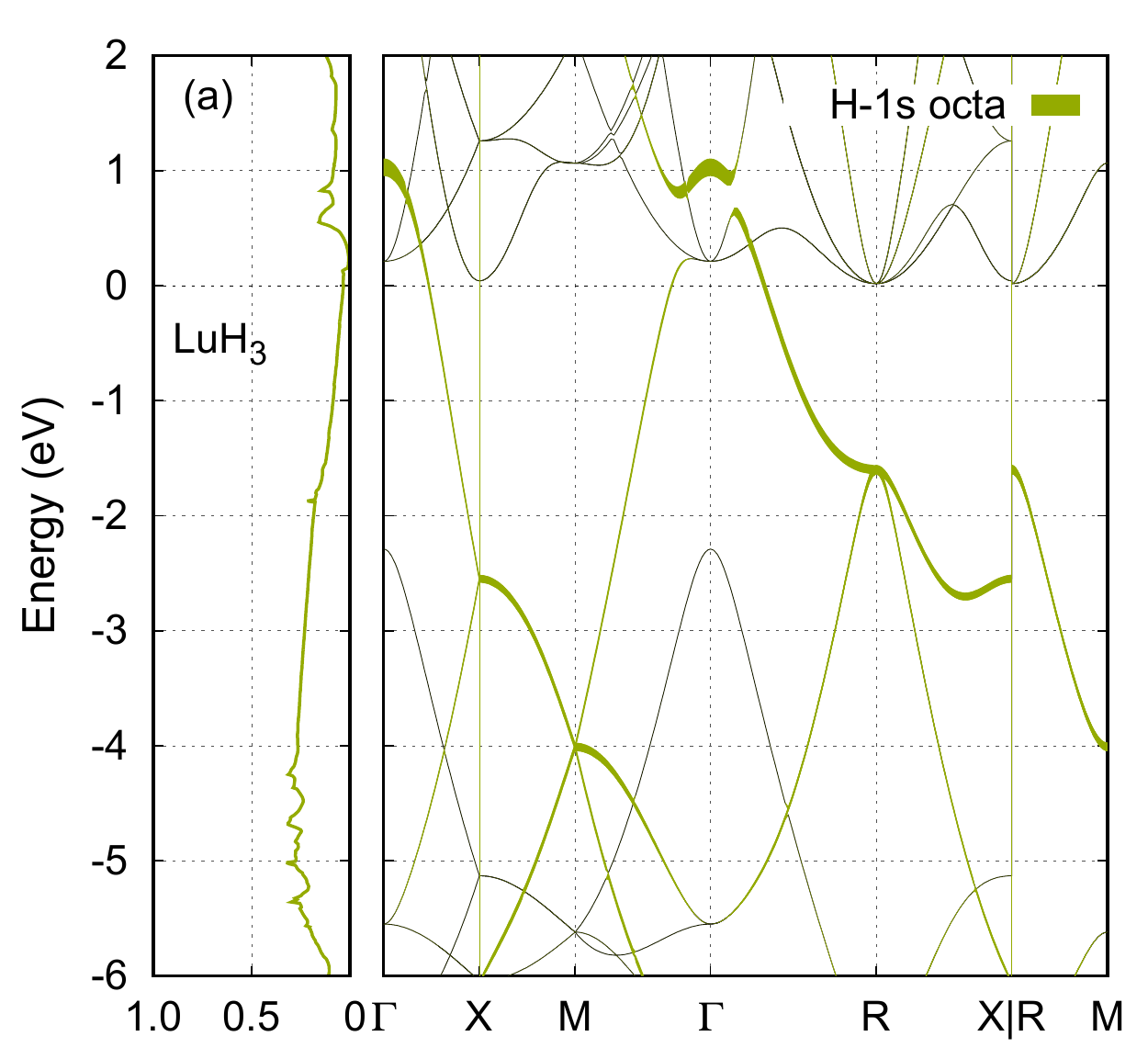}
	\includegraphics[width=0.3\linewidth]{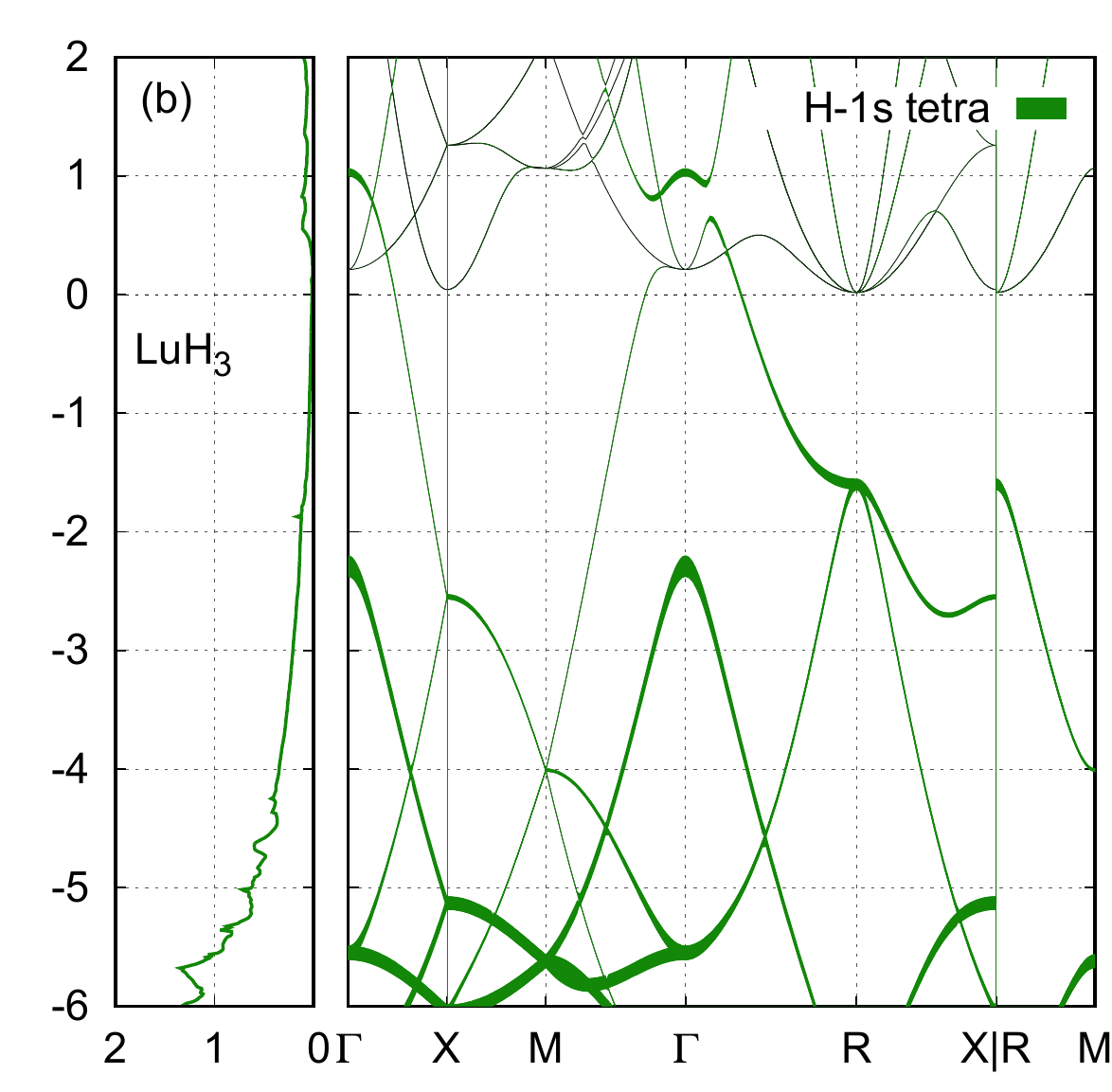}
	\includegraphics[width=0.3\linewidth]{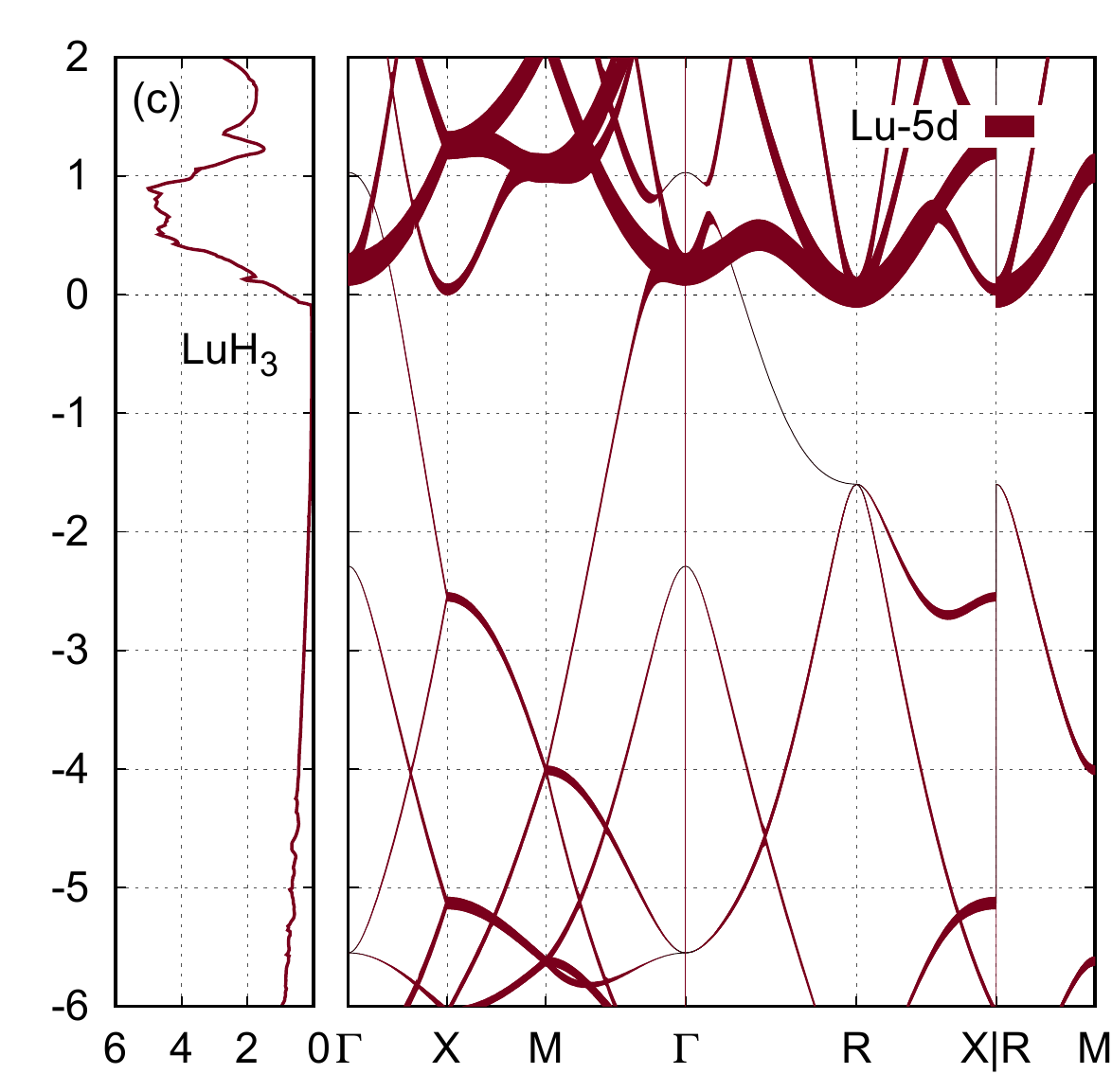}
	\includegraphics[width=0.32\linewidth]{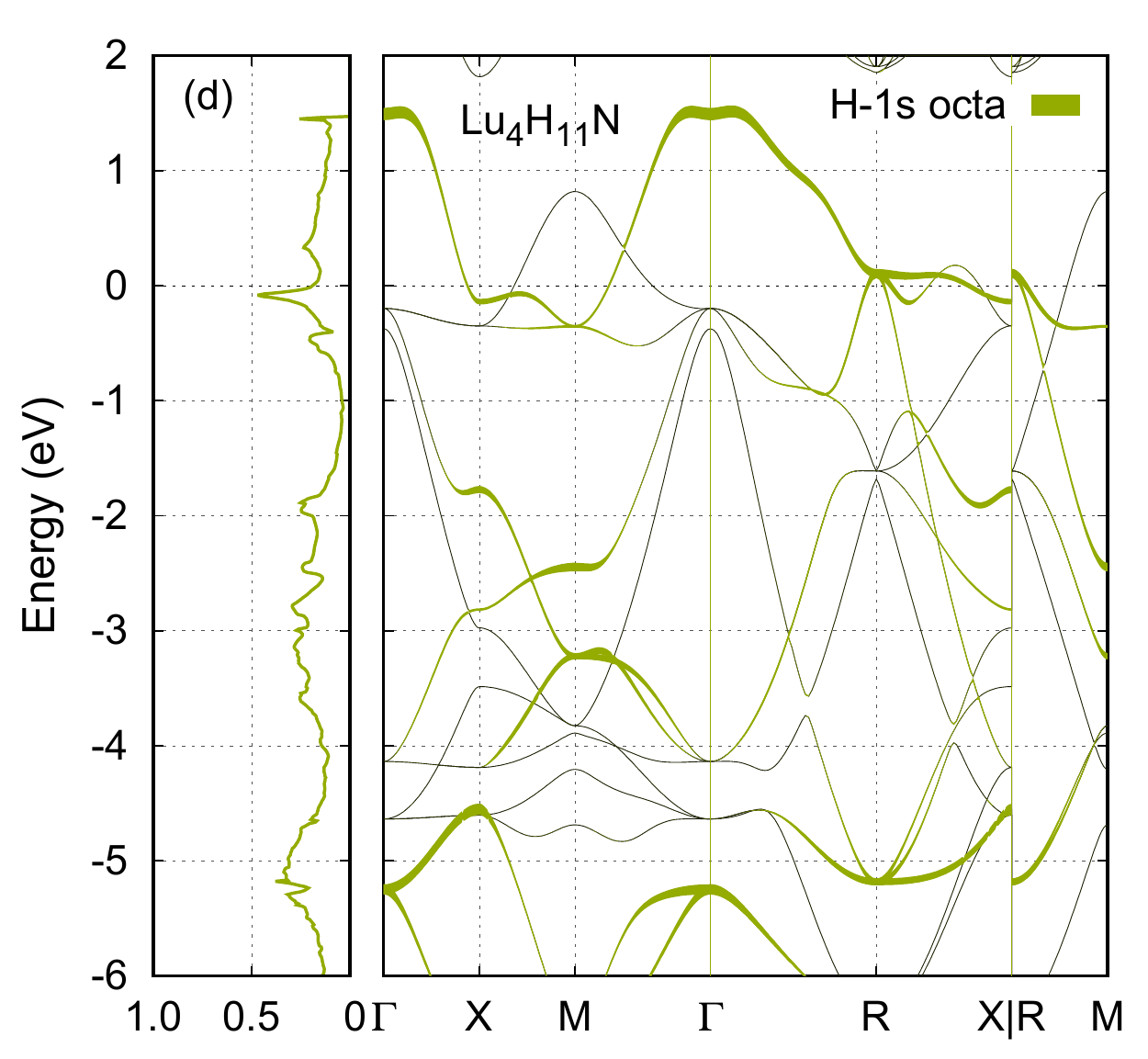}
	\includegraphics[width=0.3\linewidth]{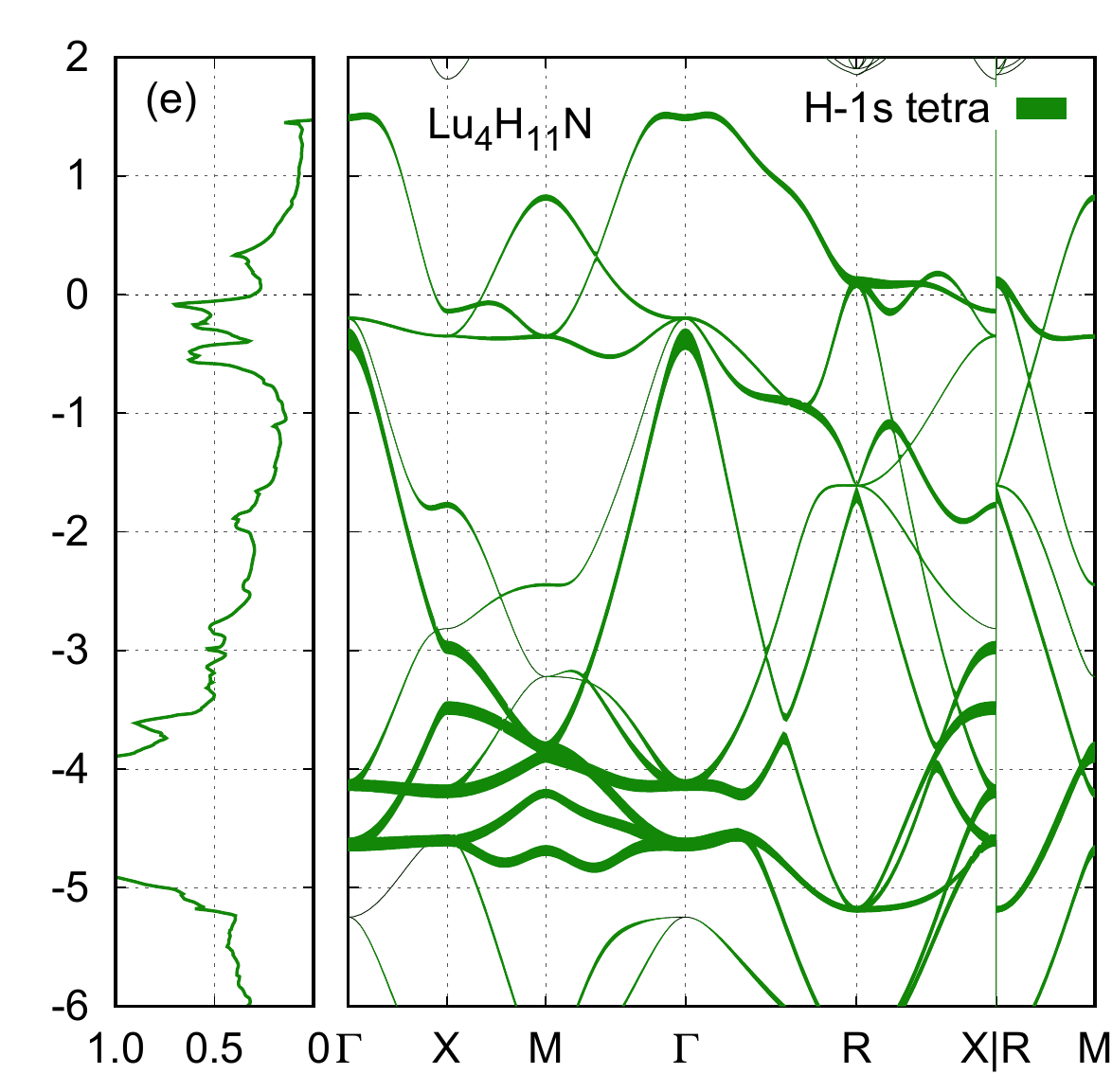}
	\includegraphics[width=0.3\linewidth]{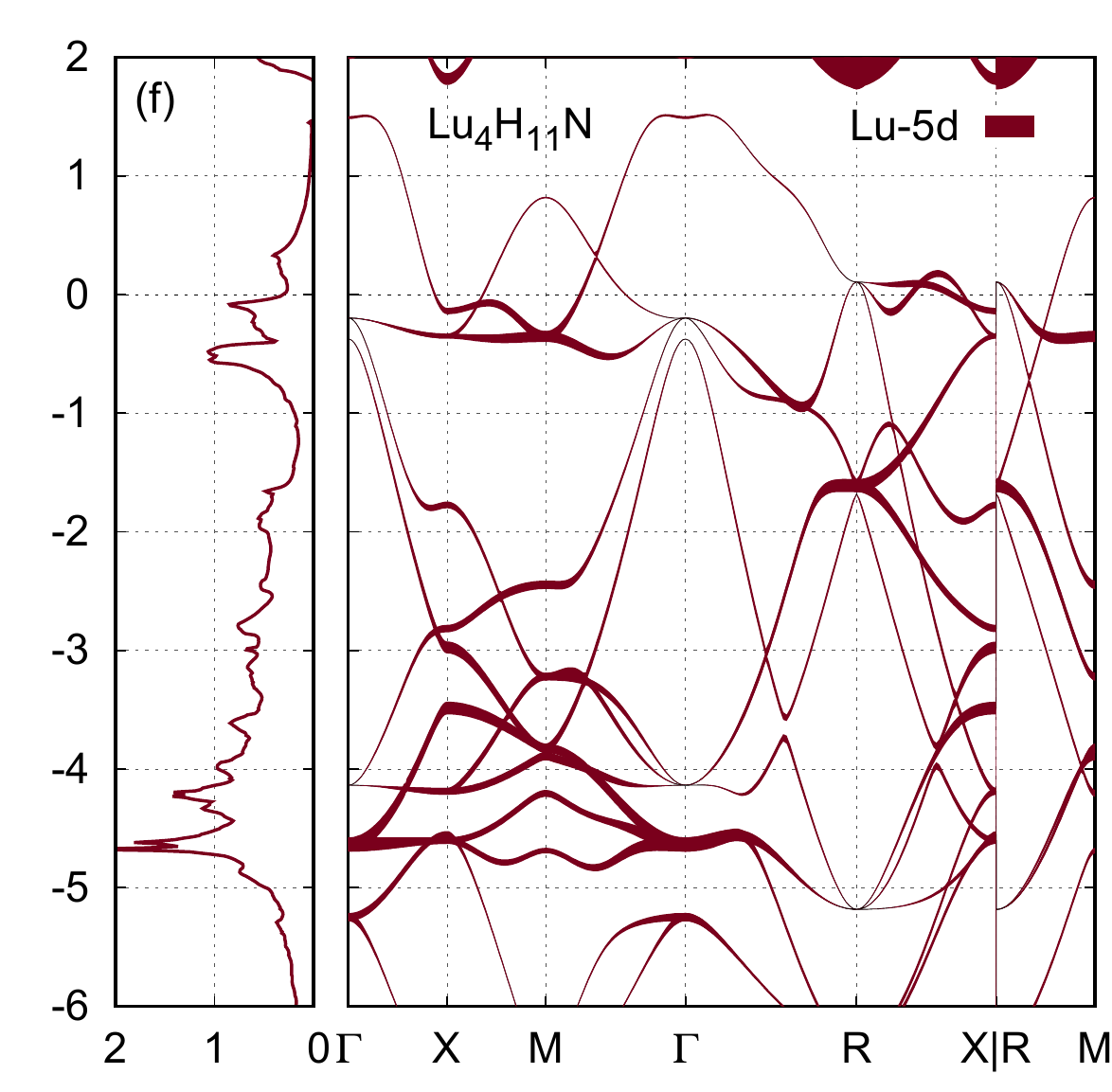}
	\includegraphics[width=0.3\linewidth]{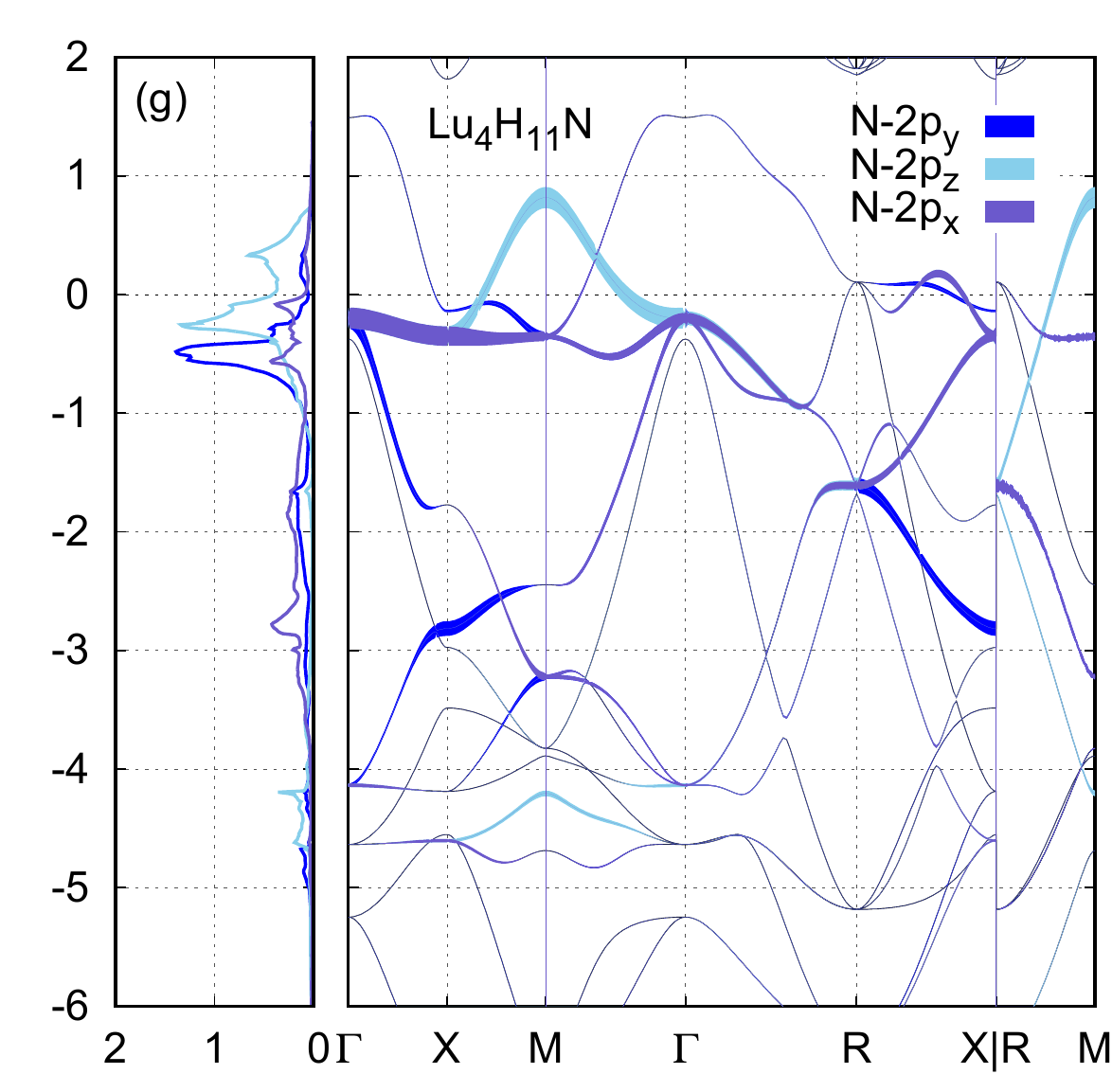}
	\caption{The orbital projected band structure, where the linewidth shows contribution of corresponding orbital: Panels (a,d) H-1s in octahedral and (b,e) tetrahedral surrounding, (c,f) Lu-5d for LuH$_{3}$ (upper row) and also (g) N-2p for LuH$_{2.75}$N$_{0.25}$ (with integer chemical indexes Lu$_{4}$H$_{11}$N) (lower row). Zero energy is the Fermi level.}
	\label{LDA_fbands}
\end{figure*}

To analyse orbital composition of bands from Fig.~\ref{LDA_DOS} let us consider contributions of orbitals with different symmetry as shown in Fig.~\ref{LDA_fbands} for LuH$_{3}$ (a-c) and LuH$_{2.75}$N$_{0.25}$ (d-g).
In case of LuH$_{3}$ main contribution to the bands above the Fermi level comes from Lu-5d states (Fig.~\ref{LDA_fbands}(c)).
The H-1s states of hydrogen atoms in octahedral surrounding manifest themselves  around 1~eV and below $-1.6$~eV (Fig.~\ref{LDA_fbands}(a)). On the other hand, the H-1s states of hydrogen atoms in tetrahedral surrounding lay predominantly below $-5$~eV (Fig.~\ref{LDA_fbands}(b)).
In contrast to LuH$_{3}$, in case of LuH$_{2.75}$N$_{0.25}$ the H-1s, Lu-5d and N-2p states contribute to the bands and density of states at the Fermi level.
Since in LuH$_{2.75}$N$_{0.25}$ there are insufficient electrons to occupy the N-2p states, they are only partially filled. Thus, N-2p bands are located around the Fermi level (Fig.~\ref{LDA_fbands}(g)). The N-2p$_z$ states cross the Fermi level, whereas the N-2p$_x$ states lay just below it. The N-2p$_y$ states are fully occupied and are located around $-1.5$~eV.
The H-1s states of hydrogen atoms in octahedral and tetrahedral surroundings lay near the Fermi level (Fig.~\ref{LDA_fbands}(d,e)).
Nevertheless, the main contribution of H-1s states of hydrogen atoms in tetrahedral surrounding is below the $-3$~eV.
At the same time the main contribution of Lu-5d states in LuH$_{2.75}$N$_{0.25}$ is located above $1.8$~eV (Fig.~\ref{LDA_fbands}(f)).

As a result, the appearance of light element H and N bands with many flat parts on the Fermi level for LuH$_{2.75}$N$_{0.25}$ compound can lead to a more favorable conditions for superconductivity than in case of the parent LuH$_{3}$ system.

\subsection{Projection to Wannier functions}
\begin{figure}[h]
		\includegraphics[width=0.85\linewidth]{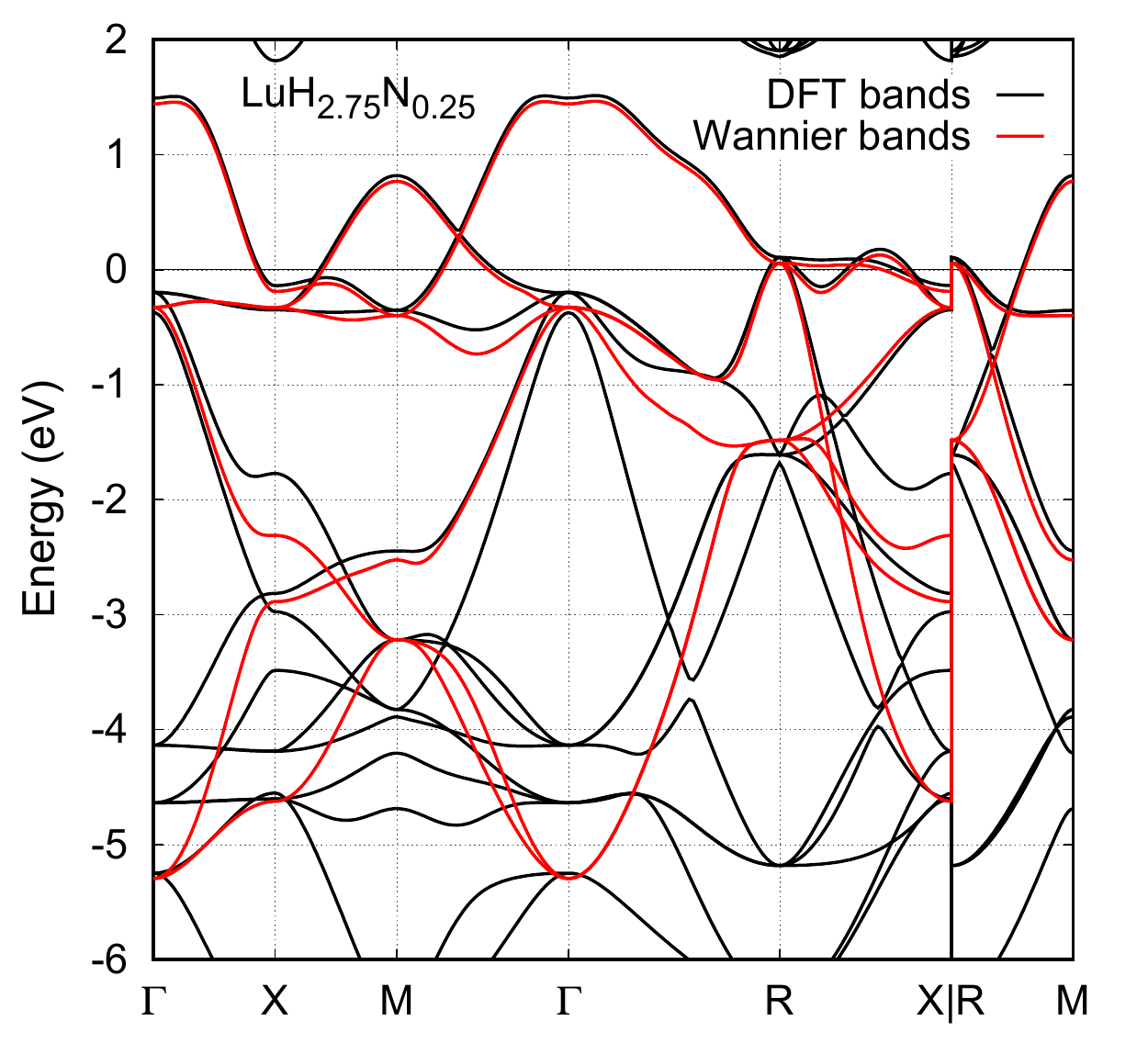}
        \includegraphics[width=0.85\linewidth]{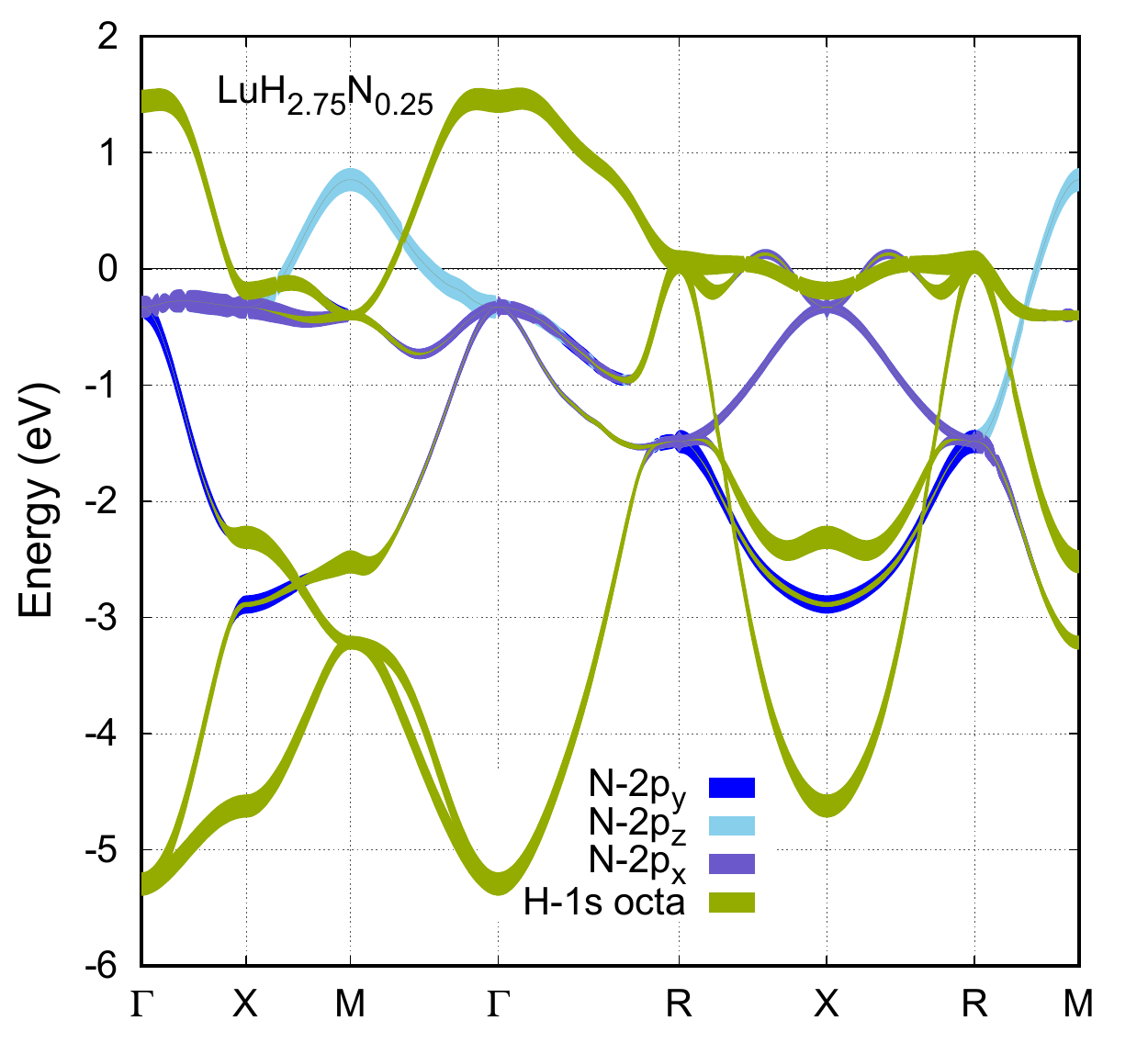}
	\caption{Comparison of original DFT bands (black) and projected on Wannier function bands (red lines) (upper). The bands projected on Wannier function  with linewidth showing contributions of H-1s octahedral and N-2p states (lower). Zero energy is the Fermi level.}
	\label{LDA_wan}
\end{figure}
In order to define the minimal orbital basis set for LuH$_{2.75}$N$_{0.25}$ to reproduce electronic bands near the Fermi level we performed the projection onto Wannier functions. Comparison of  DFT bands and bands obtained after Wannier function projection is given on Fig.~\ref{LDA_wan} (upper). It is determined that to get a good agreement between those bands one needs to include the H-1s state of three H atoms in octahedral surrounding and the N-2p states of the N atom.
The bands built on the Wannier functions coincide well enough near the Fermi level with the original DFT bands (Fig.~\ref{LDA_wan} (upper)), that indicates sufficient reliability of the projecting performed. The orbital character of the bands built on the projected Wannier functions is presented in Fig.~\ref{LDA_wan} (lower). The Projected to Wannier functions Hamiltonian are available online~\cite{Hr_online}.

\subsection{Fermi surface}
\begin{figure}[h]
		\includegraphics[width=1.0\linewidth]{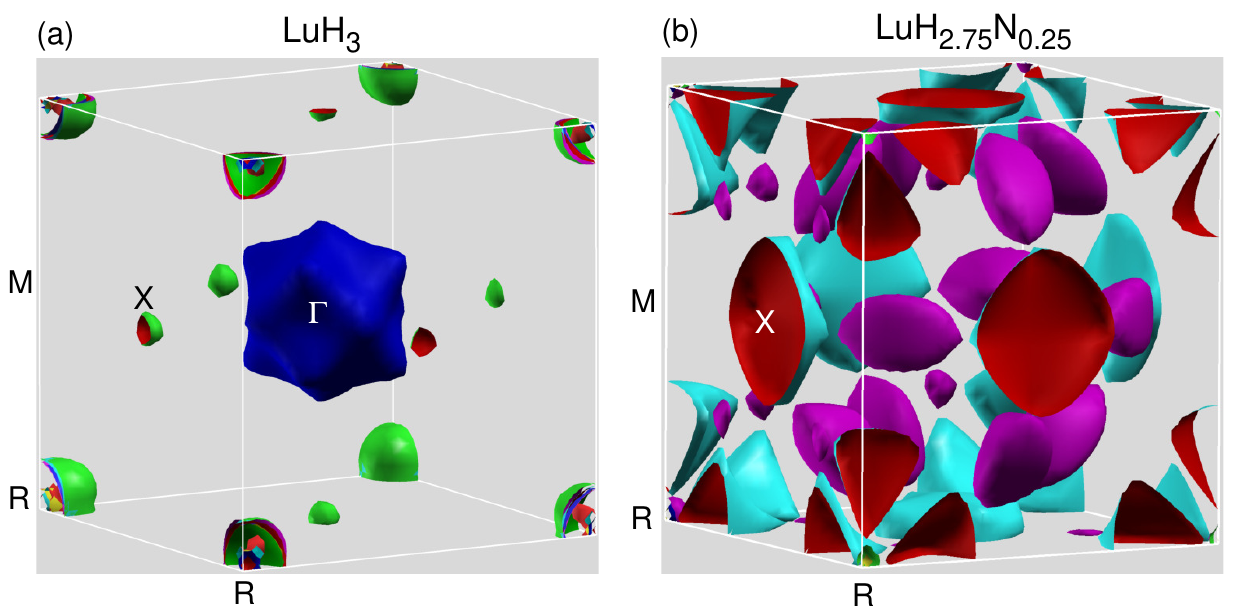}
	\caption{DFT calculated Fermi surfaces: (a) -- LuH$_{3}$, (b) -- LuH$_{2.75}$N$_{0.25}$.}
	\label{LDA_FS}
\end{figure}

The Fermi surfaces for LuH$_{3}$ (panel (a)) and LuH$_{2.75}$N$_{0.25}$ (panel (b)) are presented in Fig.~\ref{LDA_FS}.
A rather large three dimensional Fermi surface sheet of LuH$_{3}$ is seen around $\Gamma$-point. In the corners and face centers of the Brillouin zone there are some small spherical Fermi surface sheets.

For nitrogen doping case, LuH$_{2.75}$N$_{0.25}$, the Fermi surface has a more complex structure with many well developed sheets and pronounced $k$-dependence. 

\subsection{The estimation of superconducting $T_c$}
The most important parameter of superconducting materials is its critical superconducting temperature T$_c$.
To get its simplest rough theoretical estimate, we use the well known BSC equation $T_c=1.14\ \omega_D \exp(-1/\lambda)$ with Debye frequency $\omega_D$ and dimensionless pairing interaction constant $\lambda = g N(E_F)/2$ ($g$ -- pairing constant in units of energy, $N(E_F)$ -- the value of total DOS at the Fermi level). To find values of $\omega_D$ and $g$ we use the data already available in Ref.~\cite{LuH3_lucrezi_effects}.

Correspondingly, we take $T^{\rm LuH3}_{c}=62$~K as in Ref.~\cite{LuH3_lucrezi_effects} obtained with anisotropic Migdal-Eliashberg formalism. The $\omega_D=19.3$~meV $=220$~K was taken as maximal acoustic phonon energy. Once we know those values we can exclude $g$ from BCS equation and then estimate $T^{\rm LuHN}_{c} $ value for nitrogen doped material, assuming that $\omega_D$ and $g$ do not change significantly:
\begin{equation}
    T^{\rm LuHN}_{c} = 1.14\ \omega_D ( T^{\rm LuH3}_{c} / 1.14\ \omega_D )^{ N^{\rm LuH3}(E_F) / N^{\rm LuHN}(E_F) },
    \label{eq_tc}
\end{equation}
where $N^{\rm LuH3}(E_F)$ is the total DOS value at the Fermi level for LuH$_{3}$, $N^{\rm LuHN}(E_F)$ is the same but for LuH$_{2.75}$N$_{0.25}$.
Using Eq.~\ref{eq_tc} we obtained the value of $T_c\sim 111$~K for nitrogen doped lutetium hydride LuH$_{2.75}$N$_{0.25}$.

So, it can be concluded that nitrogen doping can really increase the $T_c$ value. In the case of further hole doping the density of states at the Fermi level will become higher (Fig.~\ref{LDA_DOS}). For example, the total density of states has a peak of $6.3$~states/eV/u.c. at $-0.08$~eV. With this value of the total density of states $T_c$ will be 180~K according to Eq.~\ref{eq_tc}.
It corresponds to additional 0.34 holes per unit cell with chemical composition Lu$_4$H$_{10.92}$N$_{1.08}$.
Thus it is theoretically possible to obtain relatively high $T_c$ value in the LuH$_x$N$_y$ compound. Concerning the issue of larger nitrogen doping there appears a problem of lattice stability discussed in the introduction.

\section{Conclusions}
The comprehensive investigation of the electronic structure of parent lutetium hydride LuH$_3$ and nitrogen doped lutetium hydride LuH$_{2.75}$N$_{0.25}$ are performed within the DFT+U method. The band structure, density of states, and Fermi surfaces were obtained and discussed. The Lu-5d states cross the Fermi level in the stoichiometric system LuH$_3$, while N-2p states enter the Fermi level in large quantities after nitrogen doping, bringing a significant contribution of the H-1s states. It is shown within projection onto Wannier functions, that taking into account of only the N-2p and H-1s states of hydrogen atom in the octahedral surrounding is sufficient for a good description of the bands crossing the Fermi level in case of LuH$_{2.75}$N$_{0.25}$.  

Our findings suggest that nitrogen doping of LuH$_3$ can significantly alter the electronic properties of the material (bringing light elements N-2p and H-1s states at the Fermi level), facilitating the occurrence of superconductivity. Our results also provide insights into the band structure of the nitrogen-doped lutetium hydride, which can be used to guide the design of new superconducting materials. In particular, nitrogen doping doubles the value of DOS at the Fermi level. Simple BCS analysis thus gives that for LuH$_{2.75}$N$_{0.25}$ $T_c$ the critical temperature might exceed 100~K and one can even increase it with further hole doping by nitrogen up to 180~K.

\section*{Acknowledgements}
The work of N.S.~Pavlov, K.S.~Pervakov, V.M.~Pudalov and I.A.~Nekrasov was partially supported by a grant from the Russian Science Foundation (Grant No. 21-12-00394). I.R.~Shein's work was partially supported by State Assignment No. AAAA-A19-119031890025-9. We are grateful to M.V.~Sadovskii and E.Z.~Kuchinskii for useful discussions.

\bibliography{./bib_file_LuH3}

\end{document}